\documentclass[preprint,12pt]{aastex}

\begin{document}

\title{Flux-Limited Diffusion Approximation Models of Giant Planet Formation 
by Disk Instability. II. Quadrupled Spatial Resolution}

\author{Alan P. Boss}
\affil{Earth \& Planets Laboratory, Carnegie Institution
for Science, 5241 Broad Branch Road, NW, Washington, DC 20015-1305}
\authoremail{aboss@carnegiescience.edu}

\begin{abstract}

 While collisional accumulation is nearly universally accepted as the formation
mechanism of rock and ice worlds, the situation regarding gas giant planet
formation is more nuanced. Gas accretion by solid cores formed by collisional 
accumulation is the generally favored mechanism, but observations increasingly
suggest that gas disk gravitational instability might explain the formation of at 
least the massive or wide-orbit gas giant exoplanets. This paper continues a series
aimed at refining three-dimensional (3D) hydrodynamical models of disk instabilities,
where the handling of the gas thermodynamics is a crucial factor. Boss (2017,
2019, 2021) used the $\beta$ cooling approximation (Gammie 2001) to calculate 
3D models of disks with initial masses of 0.091 $M_\odot$ extending from 4 to 20 
au around 1 $M_\odot$ protostars. Here we employ 3D flux-limited diffusion (FLD) 
approximation models of the same disks, in order to provide a superior
treatment of disk gas thermodynamics. The new models have quadrupled spatial
resolution compared to previous 3D FLD models (Boss 2008, 2012), in both
the radial and azimuthal spherical coordinates, resulting in the highest
spatial resolution 3D FLD models to date. The new models continue to support the
hypothesis that such disks can form self-gravitating, dense clumps capable of
contracting to form gas giant protoplanets, and suggest that the FLD models 
yield similar numbers of clumps as $\beta$ cooling models with $\beta \sim$ 1 
to $\sim$ 10, including the critical value of $\beta$ = 3 for fragmentation
proposed by Gammie (2001).

\end{abstract}


\section{Introduction}

 The determination that the frequency of Earth-like planets is in
the range of 0.37 to 0.88 planets per main-sequence dwarf star 
(Bryson et al. 2021) is certainly one of the most astonishing triumphs of 
modern astronomy, with consequent implications for the probability of
habitable and inhabited exoplanets throughout the galaxy. The
incredible success of the Kepler mission (Borucki et al. 2010;
Koch et al. 2010) in achieving this landmark determination has
proven that the basic scenario for the formation of our solar system,
the collisional accumulation of increasingly larger solid bodies
(Wetherill 1990, 1996), applies equally well throughout our galaxy.  
While the focus of theoretical modeling of collisional accumulation
has evolved to include novel processes such as streaming instabilities
(e.g., Youdin \& Goodman 2005; Yang et al. 2017) and pebble 
accretion (e.g., Chambers 2018; Matsumura et al. 2021), 
the basic scenario of building exoplanets from the ground up has 
remained as the leading formation mechanism.

 Within the context of the collisional accumulation scenario, the
formation of the gas and ice giant planets is hypothesized to also
proceed from the bottom up, through the accretion of protoplanetary 
disk gas by solid cores formed by collisional accumulation. Once
these cores reach masses roughly ten times that of the Earth,
Mizuno (1980) showed that their extended gaseous atmospheres 
would become unstable toward inward collapse, leading to sustained 
accretion of additional disk gas and growth to Jupiter-like masses.
There is, however, an alternative for gas giant planet formation,
the top-down mechanism of gas disk gravitational instability 
(Boss 1997), where a massive protoplanetary disk forms spiral
arms which grow and interact, resulting in the formation of
self-gravitating clumps of disk gas and dust that could contract to
form gas giant protoplanets. While disk instability is not generally
favored over core accretion (e.g., see Helled et al. 2014), 
observational evidence has suggested a role for disk instability in 
the formation of massive and wide-orbit gas giant planets, as
summarized in Boss (2017, 2019, 2021). The metallicity study
of Adibekyan (2019) similarly found a role for disk instability in
the formation of high mass exoplanets. A recent study of the 
metallicities of stars with massive planets (Mishenina et al. 2021)
found [Fe/H] ratios ranging from -0.3 to 0.4, i.e., ranging from 
metal-poor to metal-rich, compared to the solar ratio.

 Gas giant exoplanets are considerably easier to
detect than lower mass planets, but Figure 1 shows that a complete
theory of planetary system formation must take into account the
formation mechanisms of gas giants: they are not rare objects whose
existence can be ignored. In fact, a recent demographic study of Doppler 
exoplanet detections for FGKM stars (Fulton et al. 2021) has concluded 
that giant planets have an occurrence rate of about 
$14\%$ for semi-major axes of 2 au to 8 au and a rate
of about $9\%$ for semi-major axes of 8 au to 32 au. 
Given that these occurrence rates are considerably lower than
those of Earth-like planets with radii between 0.5 and 1.5 $R_\oplus$
(Bryson et al. 2021), and of the even more abundant super-Earths 
and sub-Neptunes with larger radii (Kunimoto \& Matthews 2020), it is 
surprising that Figure 1 shows so many gas giants have been detected.
The papers in this series attempt to help explain the formation
mechanism(s) of this significant exoplanet population.

 Boss (2001) calculated the first disk instability models with 3D radiative 
transfer hydrodynamics (RHD), finding fragmentation in disks that
previously had been modeled as being locally isothermal. Gammie (2001) 
hypothesized that a gas disk gravitational instability would lead to
fragmentation if the parameter $\beta$ was less than 3, where 
$\beta = t_{cool} \Omega$, with $t_{cool}$ being the disk cooling time
and $\Omega$ being the local angular velocity. 
Durisen et al. (2007) reviewed 3D RHD models where fragmentation 
did or did not occur, noting the importance of various numerical factors 
for the outcome, such as spatial grid resolution, the use of flux limiters 
for radiative transfer in the diffusion approximation, the accuracy of the 
gravitational potential solver, and the disk gas thermodynamics
controlling the heating and cooling processes. Boss (2017) discussed 
the variety of results obtained from both RHD and $\beta$ cooling 
models of disk instability, with the latter models finding critical values
for fragmentation ranging from $\beta \sim 3$ to as high as $\sim 30$,
with the critical value often depending on the spatial resolution used, 
as was also found in the 3D $\beta$ cooling simulations of Brucy \& 
Hennebelle (2021). Deng et al. (2021) found that $\beta = 3$ was
sufficient to trigger disk fragmentation in their 3D magnetohydrodynamic
models of protoplanetary disk evolution using a particle-based code.

 The present paper presents a set of FLD calculations of the same 
initial disk models considered in the FLD models of Boss (2008, 
2012) and in the $\beta$ cooling models of Boss (2017). The Boss 
(2008, 2012) models were limited in time compared to the present 
models, and more importantly, were restricted to a maximum of 
100 radial ($r$) and 512 azimuthal ($\phi$) grid points, whereas the 
new models quadrupled both of these maxima, up to 400 in radius 
and 2048 in azimuth. [The vertical resolution was unchanged, as the 
grid is compressed to the midplane in either case.] Along with the
$\beta$ cooling models of Boss (2021), these are the highest 
spatial resolution grid models ever run with the EDTONS code.
The models thus seek to approach the ideal of converging to
the continuum limit of infinite spatial resolution and to further
determine the ability of the disk instability mechanism to form
dense clumps capable of becoming gas giant planets.

\section{Numerical Methods and Flux-Limiter Derivation}

 The numerical code is the same code as that used in the many previous 
studies of gas disk gravitational instability by the author (e.g., Boss 2008, 
2012, 2017, 2019, 2021). The EDTONS code solves the three-dimensional 
equations of hydrodynamics and the Poisson equation for the gravitational 
potential, with second-order-accuracy in both space and time, on a spherical
coordinate grid (Boss \& Myhill 1992). The code uses van Leer type 
hydrodynamic fluxes and has undergone extensive testing (e.g.,
Boss \& Myhill, 1992; Boss 2007, 2009). The five hydrodynamic equations
are solved in conservation law form using explicit time differences.

 With the exception of the recent $\beta$ cooling models of Boss (2017, 2019,
2021), all of the EDTONS disk instability models since Boss (2001) have
employed radiative transfer in the diffusion approximation. This involves
solving the hydrodynamic equation which determines the evolution of the 
specific internal energy $E$:

$${\partial (\rho E) \over \partial t} + \nabla \cdot (\rho E {\bf v}) =
- p \nabla \cdot {\bf v} + \nabla \cdot \bigl[ { 4 \over 3 \kappa \rho}
\nabla ( \sigma T^4 ) \bigr], $$

\noindent
where $\rho$ is the total gas and dust mass density, $t$ is time,
${\bf v}$ is the velocity of the gas and dust (considered to be a
single fluid), $p$ is the gas pressure, $\kappa$ is the Rosseland
mean opacity, $\sigma$ is the Stefan-Boltzmann constant, 
and $T$ is the gas and dust temperature. Boss (1984) describes
in detail the equations of state used for the gas pressure, the
specific internal energy, and the dust grain opacity relation.

 Boss (2008) presented the following derivation of a flux-limiter
that could be used in concert with the EDTONS diffusion approximation 
code, which is exact only at high optical depths.  A flux-limiter can be 
employed to handle low optical depth regions (e.g., Bodenheimer et al. 
1990). The purpose of a flux-limiter is
to enforce the physical law that in low optical depth regions the ratio 
of the radiative flux ${\vec F}$ to the radiative energy density $e_r$ 
cannot exceed the speed of light $c$, i.e., $|{\vec F}| \le c e_r$.
Bodenheimer et al. (1990) adopted a prescription for enforcing
this constraint based on the flux-limiter proposed by Levermore \&
Pomraning (1981) for the situation where scattering of light is
negligible. 

 As noted by Boss (2008), the EDTONS diffusion approximation 
code is derived from a
code that handles radiation transfer in the Eddington approximation
(Boss 1984; Boss \& Myhill 1992). In this code, the energy equation is
solved along with the mean intensity equation, given by

$$ {1 \over 3} {1 \over \kappa \rho} \nabla \cdot ( {1 \over \kappa \rho} 
\nabla J) - J = -B $$

\noindent
where $J$ is the mean intensity and $B$ is the Planck function
($B = \sigma T^4 / \pi$). The mean intensity $J$ is related to the 
radiative energy density $e_r$ by $J = c e_r / 4 \pi$, while the net flux
vector ${\vec H}$ is given by ${\vec H} = {\vec F} / 4 \pi$. Hence,
the statement of physical causality $|{\vec F}| \le c e_r$ is
equivalent to $|{\vec H}| \le J$. The Eddington approximation
version of the code does not calculate ${\vec H}$ directly,
but rather $\nabla \cdot {\vec H}$, as this quantity is used
in the code to calculate the time rate of change of energy per
unit volume due to radiative transfer, $L$, through

$$ L = - 4 \pi \nabla \cdot {\vec H} = 
   {4 \pi \over 3} \nabla \cdot ({1 \over \kappa \rho} \nabla J) $$

\noindent
in optically thick regions (Boss 1984). Hence, it is convenient
to apply the physical causality constraint $|{\vec H}| \le J$
in another form. Using the equation for $L$, one finds

$$ {\vec H} = - {1 \over 3 \kappa \rho} \nabla J. $$

\noindent
The constraint $|{\vec H}| \le J$ then becomes

$$ | {4 \pi \over 3 \kappa \rho} \nabla J | \le 4 \pi J. $$

\noindent
This constraint is then evaluated in a convenient but approximate 
manner by effectively taking the divergence of both sides of this
equation, resulting in a constraint on $L$ that

$$ |L| = |{4 \pi \over 3} \nabla \cdot ({1 \over \kappa \rho} \nabla J)| 
\le |4 \pi \nabla \cdot {\vec J}|,$$

\noindent
where ${\vec J}$ is a pseudovector with $J$ as components in all
three directions. In the diffusion approximation, $J = B$.
In practice, then, $L$ is calculated for each
numerical grid point, and if $|L|$ exceeds $|4 \pi \nabla \cdot {\vec J}|$,
$L$ is set equal to $|4 \pi \nabla \cdot {\vec J}|$ but with the
original sign of $L$ (i.e., preserving the sense of whether the
grid cell is gaining or losing energy through radiative transfer).
This constraint on $L$ (Boss 2008) is thus a suitable replacement for 
the Levermore \& Pomraning (1981) flux-limiter.

\section{Numerical Spatial Resolution and Initial Conditions}

 The numerical grid has $N_r = 100$, 200, or 
400 uniformly spaced radial grid points, $N_\theta = 23$ theta grid points,
distributed from $\pi/2 \ge \theta \ge 0$ and compressed toward
the disk midplane, and $N_\phi = 512$, 1024, or 2048 uniformly spaced
azimuthal grid points. The radial grid extends from 4 to 20 au,
with disk gas flowing inside 4 au being added to the central 
protostar, whereas that reaching the outermost shell at 20 au loses
its outward radial momentum but remains on the active hydrodynamical grid.
The gravitational potential is obtained through a spherical harmonic 
expansion, including terms up to $N_{Ylm} = 48$ for all spatial resolutions.
The central protostar wobbles to preserve the center of mass of the
entire system.

 Starting with initial grid resolutions of  $N_r = 100$ and $N_\phi = 512$, 
the $r$ and $\phi$ numerical resolutions are doubled and then
quadrupled when needed to avoid violating the Jeans length 
(e.g., Boss et al. 2000) and Toomre length criteria (Nelson 2006). 
These criteria were developed to avoid spurious fragmentation
caused by inferior spatial grid resolution in Eulerian hydrodynamics
simulations of disk instability. The Jeans length criterion has also
been found recently to apply to Lagrangian hydrodynamic schemes
when studying  gravitational fragmentation (Yamamoto et al. 2021).
As in Boss (2017, 2019, 2021), if either criterion is violated,
the calculation stops, and the data from a time step prior to the 
criterion violation is used to double the spatial resolution in the 
relevant direction by dividing each cell into half while conserving 
mass and momentum. 

 Even with quadrupled spatial resolution, in some models
dense clumps formed that violated the Jeans or Toomre length
criteria at their density maxima. In that case, the calculation is
halted and restarted from a previously stored time step when
these two criteria are not yet violated. The offending high density
cell is drained of 90\% of its mass and momentum, which is then inserted 
into a virtual protoplanet (VP, Boss 2005), as in Boss (2017, 2019, 2021).
VPs are only created when a clump becomes so dense as to
violate the Jeans or Toomre length criteria and their creation
is required in order to continue the calculation with numerical rigor.
The VPs acquire circumplanetary structures that orbit along with the
VPs (e.g., Figure 2a shows two VPs embedded in bright red, high
density disk gas at between 9 o'clock and 10 o'clock for model fldA).
Such configurations are considered to be only VPs, in the sense of
the counts tabulated in Table 1; note that the clumps in Table 1
do not contain embedded VPs. The VPs orbit in the disk midplane, 
subject to the gravitational forces of the disk gas, the central protostar, 
and any other VPs, while the disk gas is subject to the gravity of the VPs. 
VPs gain mass at the rate (Boss 2005, 2013) given by the 
Bondi-Hoyle-Lyttleton (BHL) formula (e.g., Ruffert \& Arnett 1994), as
well as the angular momentum of any accreted disk gas. Any VPs that 
reach the the inner or outer boundaries are removed from the calculation. 
VPs are analogous to the sink particles used in certain Eulerian hydro 
codes, such as the FLASH (Fryxell et al. 2000) and enzo 
(https://enzo-project.org/index.html ) adaptive mesh refinement (AMR) codes.

 In the Boss (2017, 2019, 2021) models, the initial gas disk density 
distribution is that of an adiabatic, self-gravitating, thick disk with a mass
of $M_d = 0.091 M_\odot$, in near-Keplerian rotation around 
a solar mass protostar with $M_s = 1.0 M_\odot$ (Boss 1993).
The initial density distribution has a small ($1\%$) nonaxisymmetric
perturbation, while the initial temperature distribution is axisymmetric.
The initial radial temperature profile decreases from a temperature of 600 K 
at the inner boundary at 4 au to a specified initial outer disk temperature 
$T_{oi}$ at a distance of $\sim 6$ au, as shown in Figure 1 of Boss (2019).
Table 1 lists the initial temperatures in outer disks for the eight new
FLD models, which range from 40 K to 180 K, the same as the range
explored in the $\beta$ cooling models of Boss (2017). These
temperatures result in initial minimum Toomre $Q$ values for the disks
ranging from $Q_{min} = 1.3$ to $Q_{min} = 2.7$, i.e., from 
marginally unstable to relatively stable. Figure 2 of Boss (2017) shows
that the initial $Q$ value is $\sim 9.5$ at 4 au, but drops steeply to
close to the minimum value beyond $\sim$ 10 au.

 It is important to note that the disk temperatures in the Boss (2017)
$\beta$ cooling models were not allowed to decrease below their
initial values, in order to provide continuity with all of the previous
disk instability models with the EDTONS code. The same numerical
assumption was employed in the current FLD models. As noted by
Boss (2017), this means that initially high Q disks cannot become 
gravitationally unstable solely due to cooling to lower temperatures 
than that of the initial disk, and can only become more gravitationally 
unstable by transporting disk mass in such a manner that the local 
disk surface density increases, thereby lowering Q locally. 
In contrast, in the $\beta$ cooling models of Boss (2019, 2021)
the disk temperature is allowed to cool below the initial value,
to as low as 40 K. 

\section{Results}

 Figures 2 and 3 display the the final midplane density and temperature
fields for a representative sample of four of the eight models listed
in Table 1, models fldA and fldC in Figure 2, and models fldF and
fldH in Figure 3. The initial outer disk temperatures for these four
models are 40 K, 70 K, 100 K, and 180 K, respectively. As stated above, 
it is important to note that in these new models, as well as in the
$\beta$ cooling models of Boss (2017), the minimum disk temperature
at any point is defined to be the initial temperature. Hence the disks
are allowed to become hotter, but cannot cool down below their
initial temperature radial profiles. This explains the dramatically
different behavior seen in Figures 2 and 3. Model fldA, with 
an initial $Q_{min}$ = 1.3, is the most gravitationally unstable
from the start, and as a result forms the familiar pattern of spiral arms
and dense clumps in less than 200 yrs. Model fldC, somewhat more
gravitationally stable to start with an initial $Q_{min}$ = 1.7, still
manages to form well-defined spiral arms and a few clumps by
260 yrs. The situation has clearly deteriorated for the models in
Figure 3, where initial $Q_{min}$ values of 2.1 and 2.7 
characterize models fldF and fldH, respectively. Model fldF is able
to form only weak spiral arm features, whereas model fldH
settles for forming a shallow ring with no distinct clumps, as
a result of starting from a Toomre-stable initial configuration.
Note though that model fldF was able to form a single VP
during its evolution, and the VP survived in orbit through
to the final time of 287 yrs.

 Figures 2 and 3 show that the dense spiral arms are accompanied
by disk gas temperatures that rise well above the initial values as a result of
compressional heating of the disk gas to form the spiral features.
The vertical midplane optical depths in the initial models range
from $\sim 10^4$ at the inner disk boundary to  $\sim 10^2$
at the outer disk boundary, meaning that the compressional
heating cannot be quickly radiated away from the disk's surface.
The flux-limited diffusion approximation is intended to provide a
superior treatment of the proper handling of disk thermodynamics
than that provided by the classic $\beta$ cooling approximation.
Given that increasing disk temperatures produce gas pressure gradients
that resist clump formation and survival, the fact that well-defined spiral
arms and dense clumps still form in these high-spatial resolution flux-limited 
diffusion approximation models yields strong support for the hypothesis
that some gas giant protoplanets may form from a phase of
disk gas gravitational instability in a sufficiently massive and cool disk.

 Figure 4 presents a direct comparison between the compressional
heating in these FLD models and the previous $\beta$ cooling models
by showing the results for models that both started with 
$Q_{min} = 1.3$ and an outer disk temperature of 40 K.
Figure 4 displays the radial temperature and density profiles through 
a FLD model, namely model fldA's clump 3 (Table 1),  compared to a 
clump from model 1.3-3 (Table 1 in Boss 2017) with $\beta = 3$ cooling.
The present FLD models have twice the radial grid resolution as the 
Boss (2017)  $\beta$ cooling models, but in both cases the clumps at 
$\sim$ 10 au are defined by sharply peaked profiles in both the
midplane density and the temperature fields. In both cases, the vertical
optical depth at the midplane at the clump centers is greater than
$10^3$, i.e., highly optically thick. Figure 4 shows the clump in the 
$\beta$ cooling model has reached a maximum density $\sim$ 3 times
higher than the fldA clump, yet has a comparable peak temperature
of $\sim$ 100 K. This shows that the choice of $\beta = 3$ cooling
errs somewhat on the side of cooling the disk faster than the FLD 
models predict should be the case. As we shall see below, however,
while no single value of $\beta$ is able to match the FLD model results,
Figure 4 shows that a value of $\beta = 3$ does not produce grossly
disparate results compared to the comparable FLD model. A similar
result was found by Szul\'agyi et al. (2017), whose Figure 7 compares
the radial temperature gradients around clumps formed in a model
with FLD with one with local cooling, similar to $\beta$ cooling: both
treatments yielded similar clump temperature gradients, with a peak 
temperature of $\sim$ 115 K in the FLD model.

 Table 1 lists the key results for all of the models: the final times reached, 
the final spatial resolutions in $r$ and $\phi$, the final number of VPs 
and of strongly and weakly gravitationally bound clumps present at the final time,
and the sum of those two ($N_{VP}$ + $N_{clumps}$ = $N_{total}$).
As described in Boss (2021), the number of clumps
($N_{clumps}$) was assessed by searching for dense regions with
densities greater than $10^{-10}$ g cm$^{-3}$. For clumps of
this density or higher, the free fall time is 6.7 yrs or less,
considerably less than the orbital periods, implying that such clumps
might be able to survive and contract to form gaseous protoplanets,
Orbital periods of the disk gas range from 8.0 yrs at the inner edge (4 au) to 
91 yrs at the outer edge (20 au). The final times reached ranged from 189 
yrs to 1822 yrs, indicating that the models spanned time periods long 
enough for many revolutions in the inner disk and multiple 
revolutions in the outer disk. The models required over 4.5 years to 
compute, with each model running on a separate, single core of the 
Carnegie memex cluster at Stanford University. 

  Table 2 displays the clump properties for the eight new models upon 
which the numbers of  clumps in Table 1 were assessed. 
As described by Boss (2021), the clumps are classified as either 
unbound (U), weakly (W), or strongly (S) self-gravitating, depending
on whether the clump mass is less than (U), greater than (W), or more 
than 1.5 times (S) the Jeans mass for self-gravitational collapse
of a cloud with the average temperature of the clump. The Table 2
evaluations were performed at the conclusions of the calculations.
The Jeans mass is the mass of a sphere of uniform density gas
with a radius equal to the Jeans length, where the Jeans length is
the critical wavelength for self-gravitational collapse of an isothermal
gas (e.g., Boss 1997, 2005). In cgs units, the Jeans mass is given
by $1.3 \times 10^{23} (T/\mu)^{3/2}\rho^{-1/2}$, where $T$ is the gas
temperature, $\rho$ the gas density, and $\mu$ the gas mean molecular 
weight, conservatively taken here to be 2.0 for molecular hydrogen gas.
The volume-averaged clump temperature and the volume-averaged
clump density are used to compute the Jeans masses in Table 2,
conservatively producing higher Jeans masses than would be the case if
the maximum clump density was used (cf., Boley et al. 2010). The 
clumps are defined to include all cells contiguous with the maximum 
density cell with a density no more than a factor of 10 times smaller.
Clumps with masses exceeding the Jeans mass are capable of
self-gravitational collapse on a time scale given by the free fall time,
$t_{ff}$, where $t_{ff} = (3 \pi / 32 G \rho)^{1/2}$ and $G$ is the
gravitational constant. All of these clumps have a maximum densities
higher than $10^{-10}$ g cm$^{-3}$ (Table 2), implying collapse times
shorter than a free fall time of 6.7 yr, and thus considerably shorter 
times than their orbital periods. 

 The Jeans mass is of course merely an approximation for
estimating clump properties, used in part here to ensure uniformity with 
the analysis of the previous papers in this series (e.g., Boss 2019, 2021).
The Jeans mass definition is based on the analysis of the gravitational
instability of a linear wave in a uniform density gas, 
and then taking half that unstable wavelength
(the Jeans length) as being the radius of a sphere, yielding the Jeans
mass, clearly itself an approximation to a more complicated situation.
The clumps formed in these models tend to be segments of spiral
arms, often with irregular shapes, given the definition here of a clump
being all contiguous cells within a factor of 10 of the maximum density.
The heights of the clumps above the disk midplane, as defined by this
factor of 10 decrease, are comparable to their radial extents in the
disk midplane. Thus the clumps are not strongly flattened, but rather
sausage-like in cross-section. Because of their irregular shapes, a simple 
formula for their self-gravity does not exist, unlike the case of a uniform
density sphere, where the self-gravity is easily calculated 
($- 3 G M^2 / 5 R$, for mass $M$ and radius $R$). This makes
it difficult to compute the ratio of their thermal energy to their
self-gravitational energy as an alternative to the simple Jeans
mass estimates used here and in past papers (e.g., Boss 2019, 2021).
A related approach to estimating clump masses and properties was 
performed by Boley et al. (2010) for their SPH models of disk 
fragmentation at distances greater than 50 au, using both the Jeans mass
for a sphere and the Toomre mass for an unstable spiral arm.

 While it would be interesting to continue these calculations further
in time to continue to follow the evolution of the clumps and the VPs, 
that is not the purpose of this paper, which is to study the initial phases of
clump formation at unprecedented spatial resolution. Boss (2013) followed
the evolution of VPs in a gravitationally unstable disk, and found that
Jupiter-mass VPs could orbit stably for as long as $\sim$ 4000 yrs.
As an example of clump evolution in the present models, model fldB
produced a stable clump (\#1 in Tables 2 and 3) which formed and
orbited for one complete revolution by the time that the model
was terminated and the clump's properties were evaluated.
Model fldB required 1.5 yrs of CPU time before this
clump formed, and another 3 yrs to complete the final orbit, for a
total of 4.5 yrs of computer time. The final 1/4 of its orbit required
2 yrs, as that was when the highest spatial resolution was required.
Hence, in order to compute fldB's clump \#1 forward another complete
orbit at the current rate would require another 8 yrs of computation.
This estimate also ignores the fact that as the clump continues to
slowly contract and evolve, it may well violate the Jeans or Toomre
length criteria and require either the insertion of
a VP or even higher spatial resolution. The latter would further slow the 
computation. VPs are a more promising means for studying
the orbital evolution of bound clumps.

\section{Discussion}

 Figure 5 displays the masses and semi-major axes of the gravitationally 
bound clumps and VPs from Table 3, compared to the gravitationally bound 
clumps from Boss (2021), where $\beta$ cooling was used rather than 
the flux-limited diffusion approximation. Boss (2021) employed the
same high spatial resolution as the present models, and so Figure 5
contrasts the difference between the FLD approximation
and the  $\beta$ cooling approximation at an equivalent spatial
resolution. It can be seen that the clump
masses produced are similar in the two cases, but that the semi-major
axes of the FLD clumps are from $\sim$ 5 au to $\sim$ 11 au, while
those for $\beta$ cooling are from $\sim$ 8 au to $\sim$ 18 au. This 
is a result of several other differences between the two sets of models.
First, the FLD models did not allow the disk temperature to drop below
the initial radial profile, whereas the $\beta$ cooling models were allowed
to cool down to 40 K. Second, the FLD models started from initial
disks with a range in outer disk temperatures, up to 180 K for model
fldH. The $\beta$ cooling models all started from the same 
Toomre-stable outer disk temperature of 180 K, but since they were
allowed to cool down to 40 K, eventually the outer disks in the
$\beta$ cooling models became gravitationally unstable, whereas
that was prohibited for the FLD models. Comparing the final density
distributions in Figures 2 and 3 with those for the $\beta$ cooling
models in Boss (2021; Figures 2 and 3) shows that the spiral
arms in the latter models extend to the outer grid boundary, but 
are considerably more restricted to the inner disk in the present 
models. Clearly this explains the difference in semi-major axes
seen in Figure 5. Also shown for comparison are the exoplanets listed in 
the Extrasolar Planets Encyclopedia for masses between 0.1 
and 5 $M_{Jup}$ and semi-major axes between 4 au and 20 au.
Exoplanet demographics is an active and ongoing area of research,
and the present models suggest that there may be significant numbers 
of gas giants orbiting beyond 5 au, as in the models of Boss (2017,
2019, 2021), awaiting detection.

 Figure 6 again compares the results between the present FLD
models and the $\beta$ cooling models of Boss (2021). As in
Figure 5, clumps in the latter models formed at
significantly larger orbital distances, though their orbits have
eccentricities quite similar to those of the FLD models, with most
eccentricities falling below $\sim 0.15$. Clearly with quadrupled
spatial resolution, the present models, as well as those of Boss
(2021), appear to predict similar initial masses and eccentricities,
though the semi-major axes differ significantly for known reasons.

 Another comparison is to Boss (2017), where the same radial
temperature constraint was applied as in the present models.
Boss (2017) also used the $\beta$ cooling approximation for disks 
with a wide range of $\beta$ values (1 to 100), starting with initial 
disks ranging from Toomre $Q_{min} = 1.3$ to $Q_{min} = 2.7$.
The spatial resolution of Boss (2017) was limited to no more than
$N_r = 200$ and $N_\phi = 1024$, but otherwise the comparison
between the present FLD models and the $\beta$ cooling models
is appropriate. Figure 7 shows the total number of clumps and VPs 
from Table 1 for the present FLD models, compared to the total 
numbers of VPs found in the $\beta$ cooling models of Boss (2017).
Figure 7 shows that the $\beta$ = 1, 3, and 10 results span the 
results for the present FLD models, while the $\beta$ = 100  results
do not match as well. None of these four $\beta$ values matches 
the FLD results exactly. This implies a simple single value for 
$\beta$ does not do justice to representing the FLD results, 
though the inherent stochasticity of the disk gravitational instability 
mechanism makes drawing any firmer conclusion impossible.
Given this observation, Figure 7 indicates that $\beta$ cooling 
parameters in the range of $\sim$ 1 to $\sim$ 10 seem to be 
supported by the FLD models, straddling the midpoint critical
value of $\beta$ = 3 first suggested by Gammie (2001).

  Determining whether or not there is a correct critical $\beta$ value
for fragmentation continues to be controversial, as discussed by Boss 
(2017, 2019) and others. E.g., Meru \& Bate (2011b) studied convergence 
with an SPH code with as many as 16,000,000 particles, and found that 
fragmentation occurred for larger $\beta$ values as the resolution was 
increased, leading to doubt whether a critical value of $\beta$ exists. 
Meru \& Bate (2011a) found that estimates of a critical $\beta$ depend on 
the assumed underlying disk density profile and argued against the 
generality of a specific critical $\beta$ value. On the other hand,
Baehr et al. (2017) used local three-dimensional disk simulations 
to find a critical $\beta \sim 3$. Differences in the numerical model,
such as global vs. local, radiative transfer scheme, gravitational
potential solver, spatial grid resolution (as reviewed by Durisen et al. 
2007; Helled et al. 2014), and in the initial disk model,
such as the initial density and temperature distributions
(Boss 2017), make it difficult, if not impossible, for a valid comparison
of the results on critical values of $\beta$ obtained by different groups.

 Nevertheless, the Figure 7 results are similar to those of Mercer et al. 
(2018), who used two approximate radiative transfer procedures to show
that the effective value of $\beta$ could vary from $\sim$ 0.1 
to $\sim$ 200, i.e., a single, constant value of $\beta$ may not
appropriate. This suggests that more physically based cooling 
approximations than constant in space and time $\beta$ should
be investigated. E.g., Baehr \& Klahr (2015) employed a definition 
for cooling that depended on a number of factors, such as the 
midplane temperature and sound speed, the disk surface density, 
the Rosseland mean opacity, and the local angular velocity.

 Boss (2008, 2012) also calculated FLD models of disk instability.
However, the Boss (2008) study was limited to using the flux-limiter
version of the EDTONS code to continue the evolution of a
previous model that did not use a flux-limiter, in order to discern
what effect the flux-limiter might be found to have. After about
8 yr of FLD evolution, the FLD disks were found to have significantly
hotter midplane temperatures, but clump formation was still able
to proceed. Boss (2012) presented three models where the FLD
code was used from the start of the evolution, and found that clump
formation was still likely. However, these latter models were restricted
to abbreviated evolutions of only $\sim$ 65 yr and to relatively low
spatial resolution ($N_r = 100$ and $N_\phi = 512$). Thus the
present models, enabled by the computational power of the memex
cluster, provide a much more authoritative analysis of FLD
models of gas disk gravitational instability.

 Mayer et al. (2007) presented the first 3D SPH models of disk instability 
that included radiative transfer in the flux-limited diffusion approximation. 
Their initial disks were similar to those of the present model fldA, extending 
from 4 au to 20 au around a solar-mass protostar, with outer disk
temperatures of 40 K and masses of $\sim 0.1 M_\odot$. Their
Figure 1 shows the results of several disks that fragmented and
formed $\sim M_{Jup}$ clumps orbiting between $\sim$ 5 au and
$\sim$ 10 au, a result quite similar to that of the present models
(cf. Figure 5). The Mayer et al. (2007) models used $10^6$ SPH 
particles, compared to the total of $3.6864 \times 10^7$ grid cells in the
present calculations at their highest spatial resolution.

 Meru \& Bate (2010, 2011a) also presented FLD SPH models of
disk instabilities, starting with 25 au-radius disks of mass
$\sim 0.1 M_\odot$ orbiting solar-mass protostars. With 
$2.5 \times 10^5$ particles, they found that clumps formed from
$\sim$ 5 au out as far as $\sim$ 20 au, somewhat larger distances
than found in the present FLD models (Figure 5, blue points) or the 
Mayer et al. (2007) models, but similar to the clump distances obtained 
in the  $\beta$ cooling models of Boss (2021), as can also be
seen in Figure 5 (red points). The Meru \& Bate (2010, 2011a)
models started with an initial Toomre $Q$ profile that fell to a
minimum value of $\sim$ 2 at 24 au, compared to the present
models, where the minimum Toomre $Q$ value is reached
just beyond $\sim$ 10 au. Given equivalent $Q$ values, clumps 
will form faster in regions closer to the protostar, where the
orbital period is shorter, explaining the slight differences with the
Meru \& Bate (2010, 2011a) FLD results.
  
 Finally, note that the large-scale ring that formed in model fldH 
(Figure 3) did not lead to clump formation, but could still 
play an important role in gas giant planet formation. Such a
gas disk ring provides a gas pressure bump that can lead to
the rapid orbital aggregation of small solids toward the center 
of the ring, caused by gas disk headwinds for particles on orbits 
beyond that of the ring and tail winds for orbits interior to the ring's orbit
(Haghighipour \& Boss 2003). Such a pressure bump could hasten 
the pebble accretion mechanism for gas giant planet formation 
at large distances (Chambers 2021). Such a situation implies
that hybrid models of gas giant planet formation, involving both
collisional accumulation and gravitationally unstable gas disks,
should be investigated as thoroughly as the two end member 
mechanisms.

\section{Conclusions}

 The flux-limited diffusion approximation provides a formally more
accurate handling of gas disk thermodynamics, coupled with
detailed equations of state and Rosseland mean opacities, as
is the case for the present set of models, compared to
the $\beta$ cooling approximation. The spatial resolution 
in these new models, effectively equal to a hydrodynamical grid with 
as many as $3.6 \times 10^7$ spherical coordinate grid points,
is unprecedented for EDTONS FLD models, and the continued robustness
of the dense clump formation process in suitably unstable disks
supports the disk instability mechanism for gas giant protoplanet
formation as the continuum limit is approached. Compared to previous 
$\beta$ cooling models, the results suggest that $\beta \sim 3$ 
produces similar numbers of clumps as the FLD models. If such clumps 
continue to contract and survive their subsequent orbital evolution, 
these FLD models, coupled with the previous $\beta$ cooling models,
imply that significant numbers of Jupiter-mass exoplanets may orbit
at distances that are not well-sampled by ground- or space-based 
telescopic surveys to date. However, the Roman Space Telescope, scheduled
for launch in 2026, has the ability to detect these exoplanets through 
its gravitational microlensing survey. A few might even be directly 
imaged with Roman's coronagraphic instrument (CGI).

  The computations were performed on the Carnegie Institution memex computer
cluster (hpc.carnegiescience.edu) with the support of the Carnegie Scientific
Computing Committee. I thank Floyd Fayton for his invaluable assistance with 
the use of memex and the referee for several suggestions for improving
the manuscript.

\clearpage

\begin{deluxetable}{lccccccccc}
\tablecaption{Initial conditions and final results for the eight
new FLD models with up to quadrupled spatial resolution in both
the radial and azimuthal spherical coordinates compared to
Boss (2008, 2012), showing the initial outer disk temperatures
($T_{oi}$), initial minimum Toomre $Q_{min}$, the final time,
final numbers of radial and azimuthal grid points,
the number of VPs and of viable clumps at the final time,
and the sum of those two ($N_{VP}$ + $N_{clumps}$ = $N_{total}$).}
\label{tbl-1}
\tablewidth{0pt}
\tablehead{\colhead{Model} 
& \colhead{$T_{oi}$ (K)}
& \colhead{$Q_{min}$}
& \colhead{final time (yrs)} 
& \colhead{final $N_r$} 
& \colhead{final $N_\phi$} 
& \colhead{$N_{VP}$} 
& \colhead{$N_{clumps}$} 
& \colhead{$N_{total}$}}
\startdata

fldA    &  40   &  1.3  & 189.  & 400 & 2048 & 3  & 5 &  8 \\

fldB    &  60   &  1.6  & 253.  & 400 & 2048 & 2  &  1 &  3 \\
          
fldC    &  70   &  1.7  & 260.  & 400 & 2048 & 0  & 3 &  3 \\

fldD    &  80    & 1.9  & 307.  & 400 & 2048 &  0 &  3 &  3 \\

fldE    &  90    &  2.0  & 313.  & 400 & 1024 & 0  & 0 &  0 \\

fldF    &  100  &  2.1  & 287.  & 400 & 2048 & 1   & 0 &  1 \\

fldG    & 120  &  2.2  & 504.   & 200 & 1024 & 0  & 1 &  1 \\
 
fldH    & 180  &  2.7  & 1822. & 100 &  512 &  0 &  0 & 0 \\

\enddata
\end{deluxetable}

\clearpage

\begin{deluxetable}{lccccccc}
\tablecaption{Results for the new models,
showing the estimated properties for the densest clumps present at the
final times of the models, classified as either unbound (U), weakly (W),
or strongly (S) self-gravitating, based on whether the clump mass
is less than (U), greater than (W), or more than 1.5 times (S) 
the Jeans mass for self-gravitational collapse, with 
implications for clump survival. }
\label{tbl-2}
\tablewidth{0pt}
\tablehead{\colhead{Model} 
& \colhead{$Q_{min}$}
& \colhead{clump \#} 
& \colhead{$\rho_{max}$ (g cm$^{-3}$)} 
& \colhead{$M_{clump}$/$M_{Jup}$} 
& \colhead{$T_{average}$ (K)} 
& \colhead{$M_{Jeans}$/$M_{Jup}$}
& \colhead{status}}
\startdata

fldA  &  1.3  & 1 &  $1.8  \times 10^{-9}$  & 0.556  & 73.0  & 0.477  &  W  \\                              
  
fldA  &  1.3  & 2 &  $1.0 \times 10^{-9}$    & 0.699  & 53.3  & 0.498  & W   \\                              
                         
fldA  &  1.3  & 3 &  $3.9 \times 10^{-10}$    & 1.28  &  50.5 & 0.651  &  S  \\                              

fldA  &  1.3  & 4 &  $7.8  \times 10^{-10}$    & 0.845  & 53.3  & 0.542  & S   \\                              

fldA  &  1.3  & 5 &  $1.3 \times 10^{-9}$    & 0.651  & 54.4  & 0.414  &  S  \\                              

fldB  &  1.6  & 1 &  $9.47  \times 10^{-9}$    &  0.854 & 62.5  & 0.194  &  S  \\                              

fldB  &  1.6  & 2 &  $7.13  \times 10^{-10}$   & 0.946  & 124.  & 2.01  &  U  \\                              

fldB  &  1.6  & 3 &  $2.27  \times 10^{-9}$   & 1.18  &  130. & 1.24  &  U  \\                              

fldB  &  1.6  & 4 &  $6.47  \times 10^{-10}$   & 0.717  & 90.9  & 1.13  &  U  \\                              

fldC  &  1.7  & 1 &  $1.38  \times 10^{-9}$ & 1.10  & 92.5  & 0.868  & W   \\                              

fldC  &  1.7  & 2 &  $2.80  \times 10^{-9}$  & 1.21  & 74.4  & 0.482  &  S  \\                              

fldC  &  1.7  & 3 &  $1.47  \times 10^{-9}$   & 0.835  & 70.1  & 0.523  &  S  \\                              

fldD  &  1.9  & 1 &  $5.02  \times 10^{-9}$   & 0.827  & 80.0  & 0.384  &  S  \\                              

fldD  &  1.9  & 2 &  $8.53 \times 10^{-10}$   & 1.13  & 82.8  & 0.957  &  W  \\                              

fldD  &  1.9  & 3 &  $7.71  \times 10^{-10}$   & 1.90  & 81.9  & 1.05  &  S  \\                              

fldD  &  1.9  & 4 &  $3.69  \times 10^{-10}$  & 1.14  & 140.  & 3.09  &  U  \\                              

fldE  &  2.0  & 1 &  $5.05  \times 10^{-10}$  & 0.736   & 90.5  & 1.34  &  U  \\                              
   
fldF  &  2.1  & 0 &  --   & --  & --  & --  & --  \\                              
 
fldG  &  2.2  & 1 & $2.03 \times 10^{-9}$ & 3.69  & 135.  & 1.29  & S   \\                              

fldH  &  2.7  & 0 &  --  & --  & --  & --  & --   \\                              
   
\enddata
\end{deluxetable}

\clearpage

\begin{deluxetable}{lccccccc}
\tablecaption{Masses and orbital parameters for the gravitationally bound
clumps, i.e., those with a status of W or S in Table 2, as well as
for the VPs, both at the final times listed in Table 1.}
\label{tbl-3}
\tablewidth{0pt}
\tablehead{\colhead{Model} 
& \colhead{$Q_{min}$}
& \colhead{clump \#} 
& \colhead{VP \#} 
& \colhead{$M/M_{Jup}$} 
& \colhead{semimajor axis (au)}
& \colhead{eccentricity}
& \colhead{status}}
\startdata

fldA  &  1.3  & 1 & -- & 0.556  & 8.62  & 0.101  &  W  \\                              
  
fldA  &  1.3  & 2 & -- &  0.699 & 10.35  & 0.111  &  W  \\                              
                         
fldA  &  1.3  & 3 & -- &  1.28  & 10.37  & 0.037  &  S  \\                              
  
fldA  &  1.3  & 4 & -- &  0.845 & 10.91  & 0.053  &   S  \\                              

fldA  &  1.3  & 5 & -- &  0.651 & 8.23  & 0.117  &  S  \\                              
  
fldA  &  1.3  & -- & 1 & 0.31  & 8.34  & 0.057  &  --  \\                              
                           
fldA  &  1.3  & -- & 2 &  0.023 & 7.38  & 0.140  &  -- \\                              
  
fldA  &  1.3  & -- & 3 &  0.92 & 5.42  & 0.123  &  -- \\                              
                           
fldB  &  1.6  & 1 & -- &  0.854 & 9.98  & 0.141  &   S \\                                                     

fldB  &  1.6  & -- & 1 & 0.515 & 4.73   & 0.505  &  -- \\                              

fldB  &  1.6  & -- & 2 &  0.540 &  --  & --  &   unbound  \\                              

fldC  &  1.7  & 1 & -- &  1.10 & 8.86  & 0.136  &   W \\                              

fldC  &  1.7  & 2 & -- &  1.21 & 7.44  & 0.0558  &  S \\                              

fldC  &  1.7  & 3 & -- &  0.835 & 9.56  & 0.108  &   S \\                              

fldD  &  1.9  & 1 &  -- & 0.827  & 8.49 &  0.0698 &   S  \\                              

fldD  &  1.9  & 2 &  -- & 1.13 & 10.5 & 0.0514  &   W  \\                              

fldD  &  1.9  & 3 &  -- & 1.90 & 10.7  & 0.0813  &   S \\                              

fldE  &  2.0  & -- &  -- & --  & -- & -- & -- \\                              
   
fldF  &  2.1  & -- &  1 & 0.750 & 5.06 & 0.154  &   -- \\                              
 
fldG  &  2.2  & 1 & -- &  3.69 & 8.58  & 0.155 & S \\  
                      
fldH  &  2.7  & --  & -- & --  & -- & -- & -- \\  
  
\enddata
\end{deluxetable}

\clearpage

\begin{figure}
\vspace{-3.0in}
\plotone{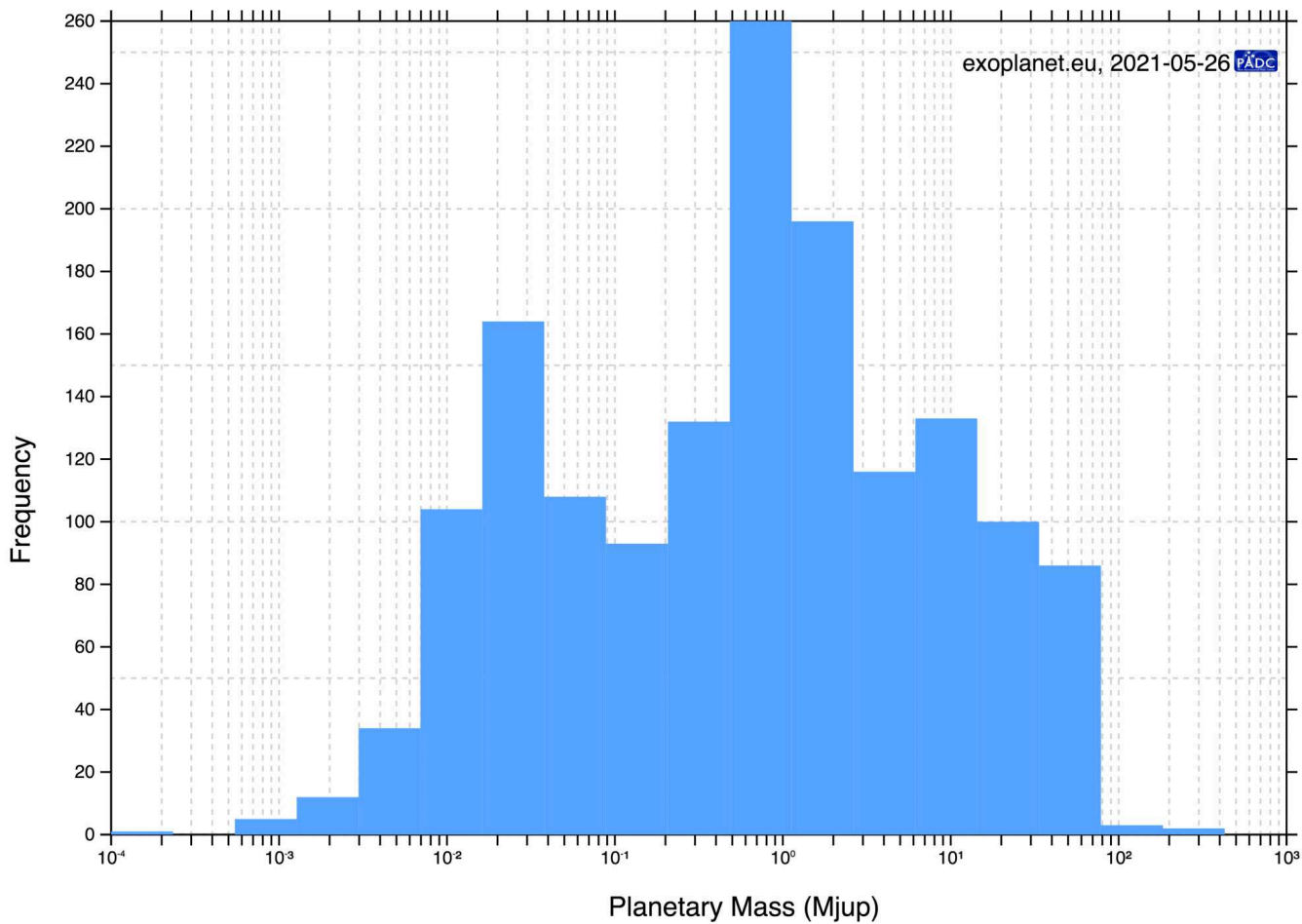}
\vspace{-1.0in}
\caption{Frequency of planets in the Extrasolar Planets Encyclopedia database 
(exoplanet.eu) as a function of planetary mass for confirmed mass planets ranging 
between $10^{-4} M_{Jup}$ and $10^3 M_{Jup}$. Super-Earths and sub-Neptunes 
with masses of $\sim 2 \times 10^{-2} M_{Jup}$ represent one peak in the distribution,
while Jupiter-mass gas giants represent an even higher peak, largely as a result of
the relative ease of their detection. }
\end{figure}

\clearpage

\begin{figure}
\vspace{-2.0in}
\plotone{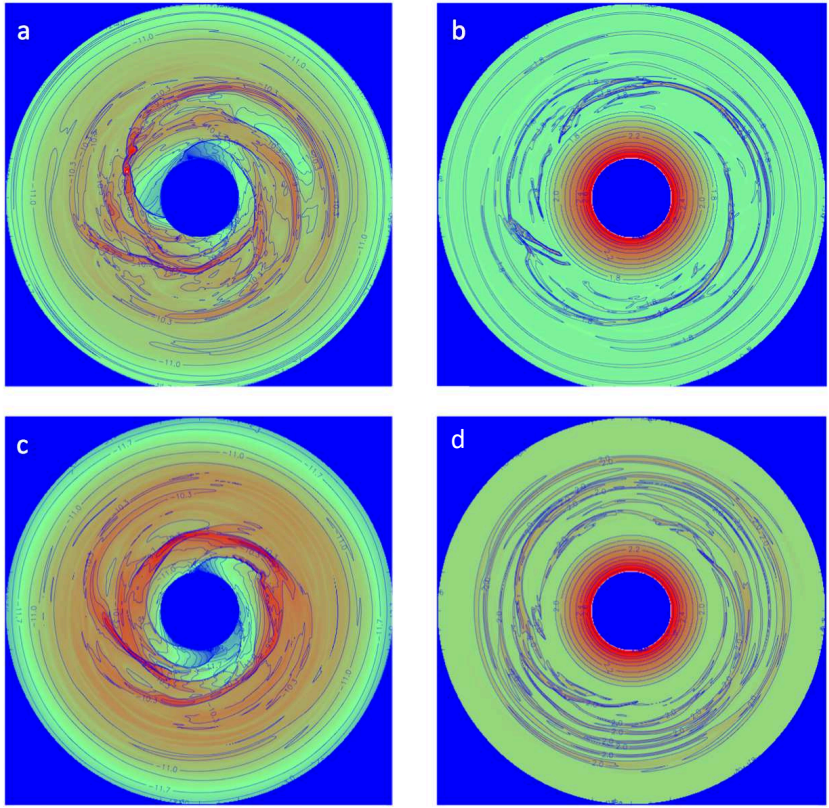}
\vspace{-1.0in}
\caption{Equatorial (midplane) density (left) and temperature (right) contours for 
models fldA (a,b) at 189 yrs and for model fldC (c,d) at 260 yrs, respectively.
The disks have inner radii of 4 au and outer radii of 20 au. Density contours are labelled in 
log cgs units and temperature contours are labelled in log K units.
Maximum midplane gas densities are:
(a) $1.35 \times 10^{-8}$ g cm$^{-3}$ and  (c) $3.31 \times 10^{-9}$ g cm$^{-3}$.
The initial maximum midplane density is $1.0 \times 10^{-10}$ g cm$^{-3}$ at 4 au.
The models start with initial outer disk temperatures ranging from 40 K (light green color)
for model fldA to 180 K (light orange color) for model fldH. Red colors 
correspond to temperatures of $\sim$ 600 K close to the inner boundary.}
\end{figure}

\clearpage

\begin{figure}
\vspace{-2.0in}
\plotone{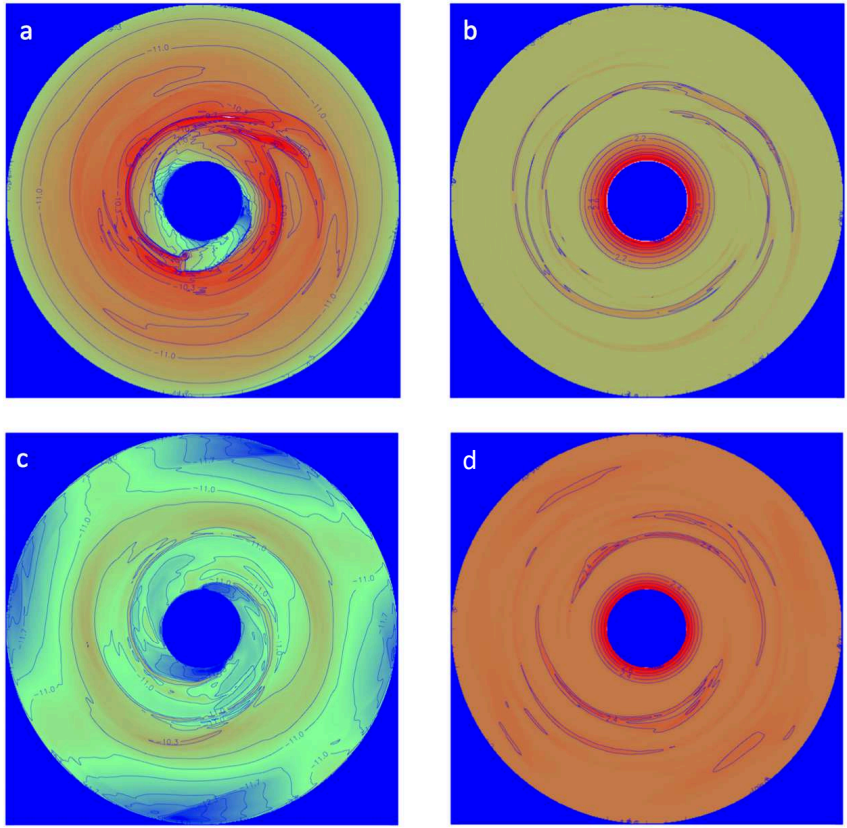}
\vspace{-1.0in}
\caption{Equatorial (midplane) density (left) and temperature (right) contours for 
models fldF (a,b) at 287 yrs and for model fldH (c,d) at 1822 yrs, respectively,
plotted as in Figure 2. Maximum midplane gas densities are:
(a) $8.71 \times 10^{-10}$ g cm$^{-3}$ and  (c) $5.01 \times 10^{-11}$ g cm$^{-3}$.}
\end{figure}

\clearpage

\begin{figure}
\vspace{-2.0in}
\plotone{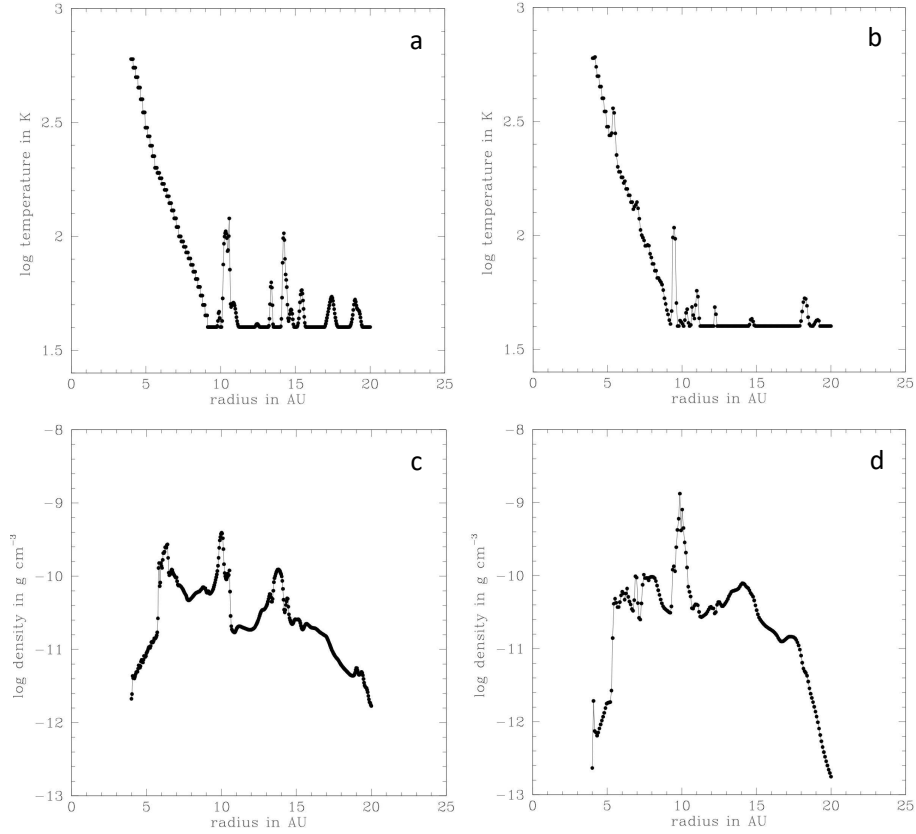}
\caption{Radial temperature (a) and density (c) profiles through a representative
FLD model highlighting the model fldA clump 3 (Table 1), seen at about 3 o'clock 
in Figure 2 at an orbital distance of $\sim$ 10 au, compared to a representative
clump (b, d) at a similar time of evolution (174 yr)
from model 1.3-3 (Table 1 in Boss 2017), where $\beta = 3$ cooling 
was used rather than the flux-limited diffusion approximation. In both cases the 
clumps orbit at $\sim$ 10 au and are defined by sharply peaked profiles in both the
midplane density and the temperature fields.}
\end{figure}

\clearpage

\begin{figure}
\vspace{-2.0in}
\plotone{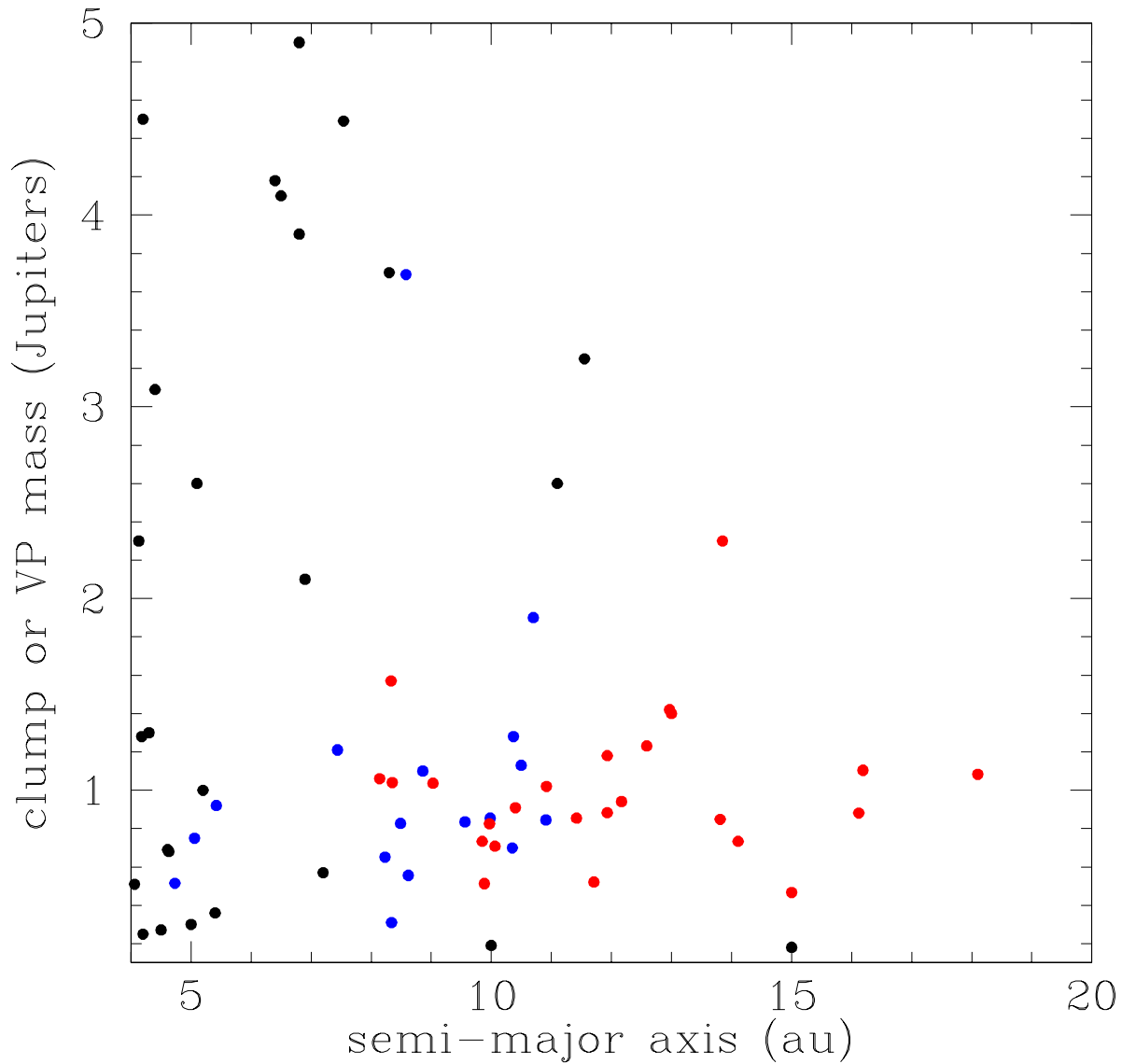}
\caption{Masses and semi-major axes of the gravitationally bound clumps 
and VPs from Table 3 are shown in blue, while the red points show the 
gravitationally bound clumps from Boss (2021), where $\beta$ cooling was
used rather than the flux-limited diffusion approximation. The black points
show all exoplanets listed in the Extrasolar Planets Encyclopedia 
(exoplanets.eu) for masses between 0.1 and 
5 $M_{Jup}$  and semi-major axes between 4 au and 20 au.}
\end{figure}

\clearpage

\begin{figure}
\vspace{-2.0in}
\plotone{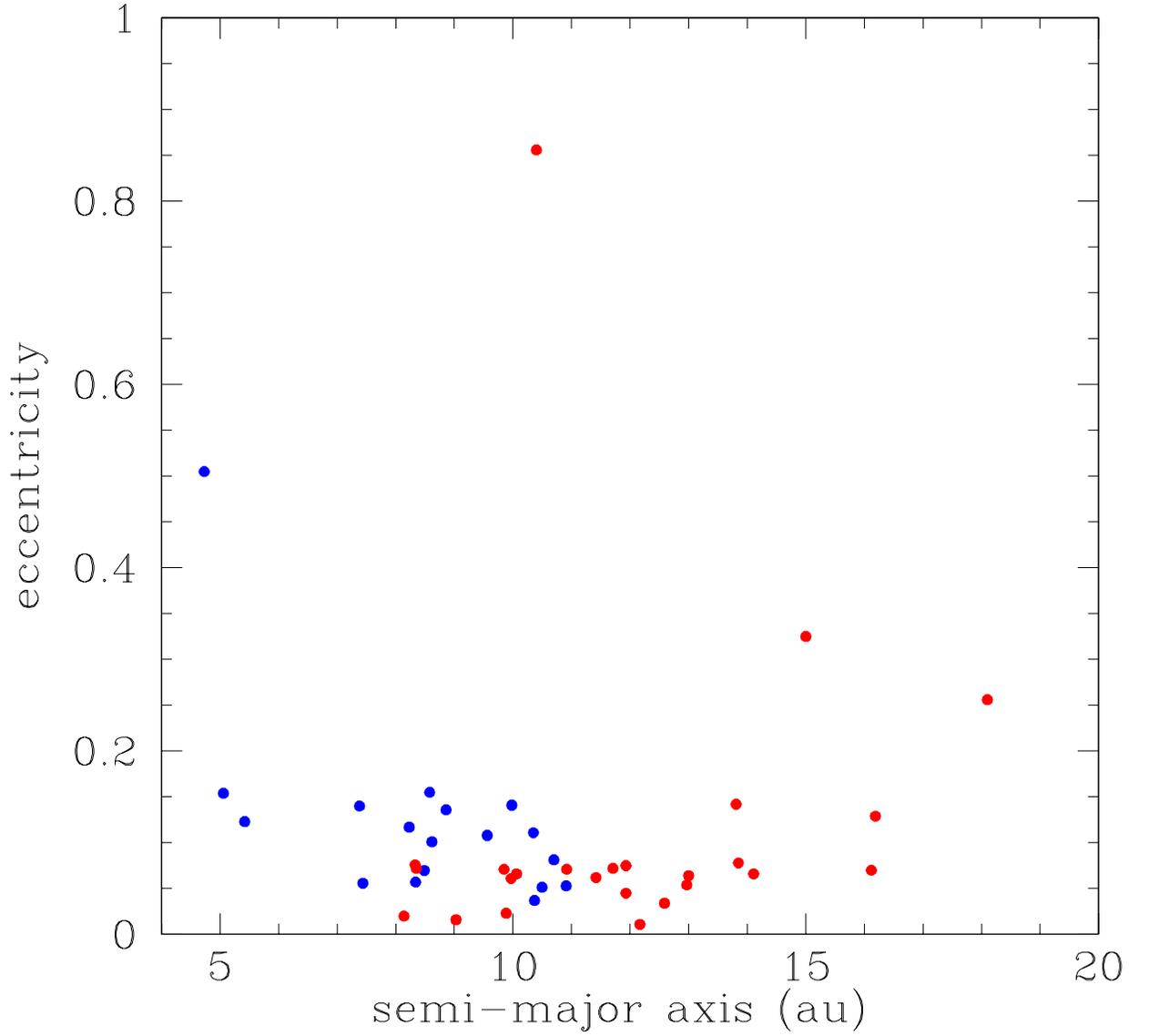}
\caption{Eccentricities and semi-major axes for the gravitationally bound clumps 
and VPs listed in Table 3 are shown in blue, while the red points show the
gravitationally bound clumps from Boss (2021), where $\beta$ cooling was
used rather than the flux-limited diffusion approximation.}
\end{figure}

\clearpage

\begin{figure}
\vspace{-2.0in}
\plotone{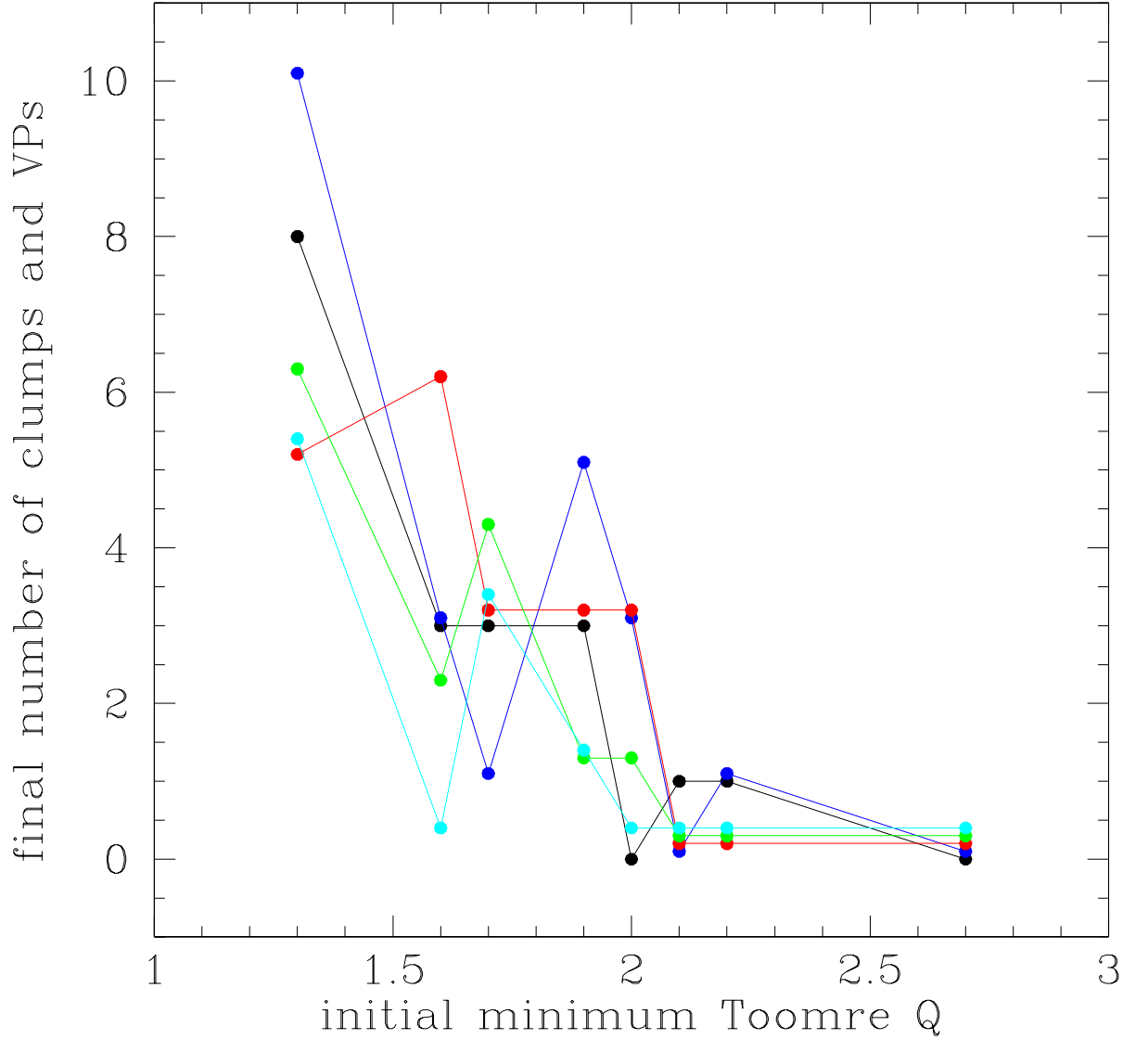}
\caption{Black points and connecting lines plot the total number of clumps 
and VPs from Table 1 for the present eight FLD models, each starting with a
different initial outer disk temperature and hence with a different initial 
minimum $Q$, while the four other colors represent the total numbers of 
VPs found in the equivalent initial minimum $Q$ $\beta$ cooling model of
Boss (2017), with blue for $\beta$ = 1, red for $\beta$ = 3,  green for 
$\beta$ = 10, and cyan for $\beta$ = 100. The VP numbers for the colored
points are adjusted upwards by 0.1 for each increase in $\beta$ for clarity.}
\end{figure}

\clearpage

\end{document}